\documentclass[aps,prl,twocolumn,groupedaddress]{revtex4}

\usepackage{color}
\usepackage{graphicx}
\usepackage{amssymb, amsmath, amsfonts}

\begin{document}


\title{Light-induced structural transformations in a single gallium nanoparticulate}


\author{B. F. Soares}
\author{K. F. MacDonald}
\author{V. A. Fedotov}
\author{N. I. Zheludev}
\affiliation{EPSRC NanoPhotonics Portfolio Centre, School of
Physics and Astronomy, University of Southampton, SO17 1BJ, UK\\}


\date{\today}

\begin{abstract}
In a single gallium nanoparticulate, self-assembled (from an
atomic beam) in a nano-aperture at the tip of a tapered optical
fiber, we have observed evidence for a sequence of reversible
light-induced transformations between five different structural
phases ($\gamma \rightarrow \varepsilon \rightarrow \delta
\rightarrow \beta \rightarrow liquid$), stimulated by optical
excitation at nanowatt power levels.
\end{abstract}

\pacs{64.70.Nd, 78.67.Bf}

\maketitle


In this letter we report experimental evidence for a sequence of
light-induced structural phase transformations involving five
different phases in a single gallium nanoparticulate. Our results
are consistent with predictions that phase transitions in
nano-volumes of material are achieved through \emph{continuous}
and \emph{reversible} surface-driven coexistences of different
forms \cite{Berry00-JCP113, Shirinyan04-Nano15}, and demonstrate
that such transformations can be stimulated and controlled by
extremely low power optical excitation. We have been able to
induce and monitor transitions between phases that differ in free
energy by only a fraction of an $meV$ per atom and found that the
nanoparticulate's structural response to optical excitation
settles within a few tens of microseconds. We observed that a
particulate probed with low intensity cw laser light can be
overcooled by more than $90$ $K$ before it returns to the
low-temperature phase, but that under a regime of more intense
pulsed optical excitation this overcooling hysteresis collapses.

With such techniques as confocal microscopy, `optical tweezers'
and scanning near-field imaging it is now possible to detect
photoluminescence and Raman spectra and perform sophisticated
transient spectroscopic measurements on single nanoparticles
\cite{Prikulis04-NL4, Lindfors04-PRL93, Peyser01-Science219,
Doering02-JPCB106} - essentially removing the inhomogeneous
broadening characteristic of nanoparticle film spectroscopy. In
nanoparticles, phase transition temperatures and optical
absorption cross-sections depend strongly on particle size, so
inhomogeneous broadening in nanoparticle films masks the
characteristic changes in optical properties that accompany a
transition. Studying transitions in a single nanoparticulate, as
opposed to a nanoparticle film \cite{MacDonald03-APL82,
Pochon04-PRL92}, leads to advantages similar to the resolution
improvements achieved in the optical spectroscopy of single
particles.

\begin{figure}
\includegraphics[width=85mm]{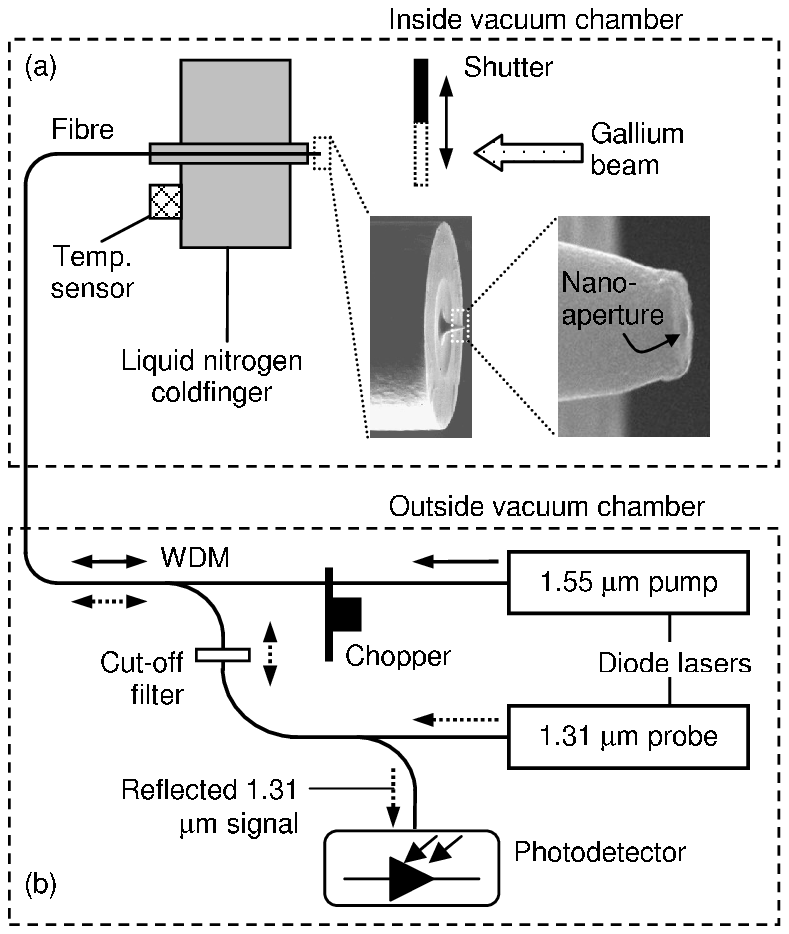}
\caption{Experimental arrangement for studying light-induced
structural transformations in a single gallium nanoparticulate.
(a) Components inside the UHV chamber, and SEM images of the
tapered fiber tip. (b) Fiber-optic arrangement outside the chamber
for reflective pump-probe measurements.  \label{Fig1}}
\end{figure}

\begin{figure*}
\includegraphics{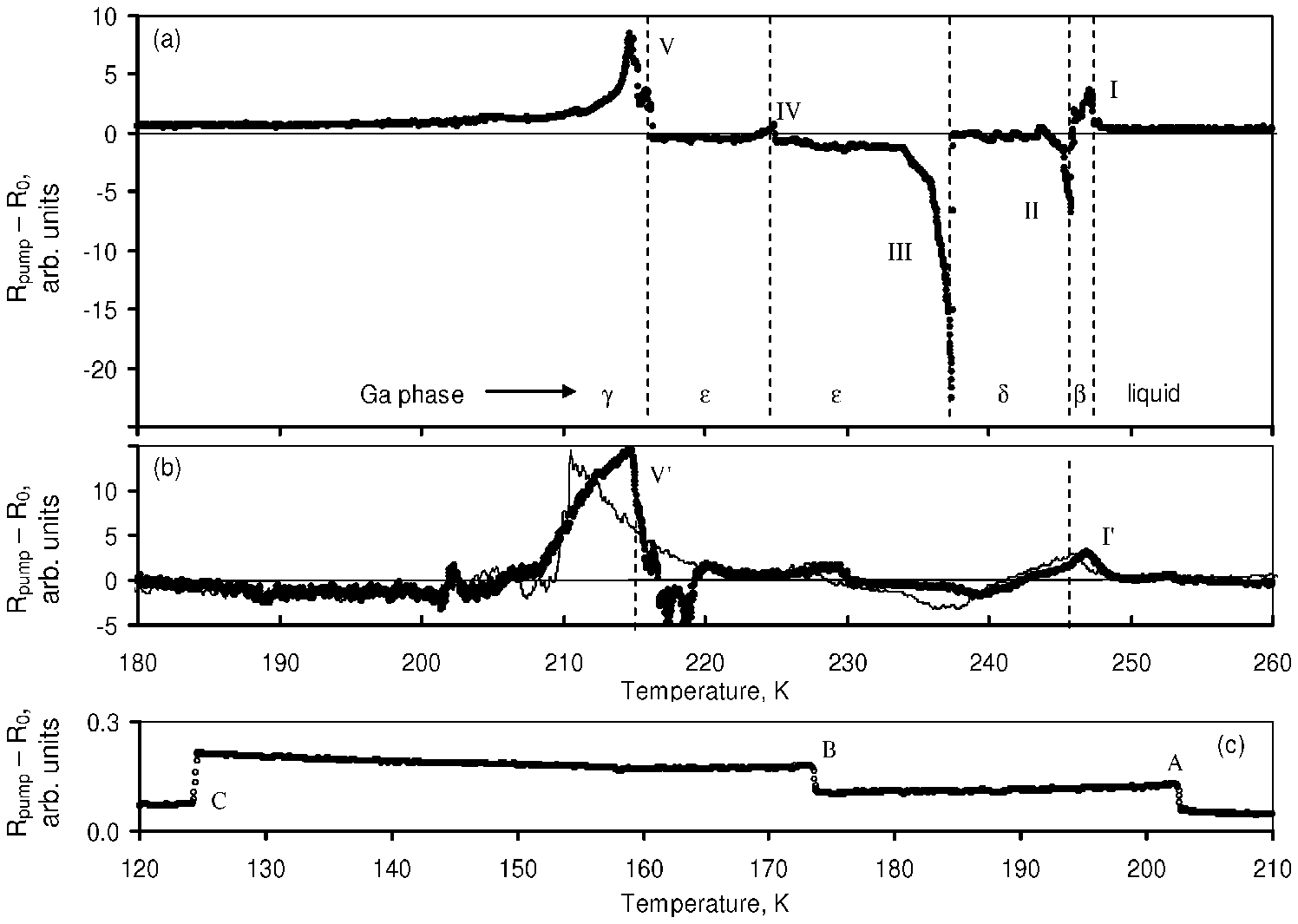}
\caption{(a) Magnitude ($R_{pump} - R_{0}$) of pump-induced
reflectivity change for a gallium nanoparticulate as a function of
increasing temperature ($R_{pump} =$ reflectivity with pump
illumination, $R_{0} =$ reflectivity with no pump present. cw
probe power at particulate $\sim 20$ $nW$, pump peak power $\sim
30$ $nW$, plotted temp. = measured temp.). (b) $R_{pump} - R_{0}$
(increasing and decreasing temperature) for the same particulate
but with $3$ $ns$ pump pulses (rep. rate $30$ $kHz$, peak power
$100$ $mW$, plotted temp. = measured temp. + 65) (c) Decreasing
temperature part of the temperature cycle shown in (a).  Note that
the recorded temperature of the coldfinger is somewhat lower than
the actual temperature of the nanoparticulate due to localized
laser-heating.\label{Fig2}}
\end{figure*}

We studied a single nanoparticulate, grown from an atomic beam, at
the tip of a tapered optical fiber with a nano-aperture at its
end. This location allows for precise coupling of optical
excitations to the particulate for the stimulation of phase
transformations, and simultaneously for collection of a probe
signal reflected by the particulate and used to monitor its state.
We used a gold-coated silica single-mode fiber tapered to a $100$
$nm$ aperture (see inset to Fig.~\ref{Fig1}a). It was attached to
a liquid nitrogen cryostat cold-finger at $80$ $K$ inside a vacuum
chamber evacuated to $\sim 10^{-6}$ $mbar$. To grow the
nanoparticulate, a gallium atomic beam with a mass flux of $f \sim
0.3$ $nm/min$ was directed at the end of the fiber for $t = 60$
$min$ (Fig.~\ref{Fig1}a). The formation of a particulate in the
nano-aperture was indicated by a change in the reflectivity of the
fiber tip during deposition (we assume that, at least on
formation, the particulate fills the aperture and therefore has a
diameter equal to that of the aperture, i.e. $100$ $nm$). A cw
diode laser operating at $1310$ $nm$ was used as a probe and
another at $1550$ $nm$, typically modulated at $\nu=2.3$ $kHz$,
was used as the pump. The reflected probe signal was monitored
using an InGaAs photodetector and a digital lock-in amplifier
(locked on the frequency $\nu$). A wavelength-division multiplexer
(WDM) and bandpass filter prevented reflected pump light from
reaching the photodetector (see Fig.~\ref{Fig1}b). Structural
transformations were observed by monitoring
\emph{pump-beam-induced changes} in the reflectivity of the
particulate as a function of temperature between $80$ and $300$
$K$ (varied at a rate of $\sim 2$ $K/min$) using pump and probe
powers of $30$ and $20$ $nW$ respectively at the nano-aperture.
While increasing the temperature of the nanoparticulate we observe
several narrow peaks in the induced-reflectivity-change signal at
temperatures between $200$ and $250$ $K$, as shown in
Fig.~\ref{Fig2}a (no features are observed outside the range
shown). Positive and negative peaks correspond to
\emph{pump-induced} increases and decreases in nanoparticulate
reflectivity. When repeating the heating cycle the peaks appear at
the same positions, although their relative heights may vary
slightly.

Substantial changes in the optical properties of the
nanoparticulate may occur when it undergoes a transition between
two structural phases. Such changes are typically much more
dramatic than temperature-dependent variations occurring within a
single phase. Indeed, in the case of gallium, differences between
the electronic density of states of its various phases
\cite{Bernasconi95-PRB52} lead to pronounced differences between
their optical dielectric coefficients. A phase change in a galluim
nanoparticulate thus affects its optical absorption cross-section
\cite{Fedotov04-JOA6}, and in the present case will change the
reflectivity of the nano-aperture hosting the particulate. One may
thus detect phase transformations in a nanoparticulate by
monitoring its optical properties. A peculiarity of confined
solids is that their structural transformations take the form of a
dynamic coexistence between different phases (as opposed to an
abrupt transition) \cite{Berry00-JCP113, Shirinyan04-Nano15}, with
the surface of a particle (where atoms have fewer nearest
neighbors than internal atoms) acting as a boundary at which
transformations start \cite{Peters97-APL71, Parravicini03-APL82}.
In the presence of optical excitation, the phase equilibrium (and
therefore in the present case the reflectivity of the particle) is
determined by both temperature and the level of electronic
excitation \cite{Wautelet04-JPCM16}. In the current experiment,
incident laser light leads to both electronic excitation and a
local temperature increase, which simultaneously affect the phase
equilibrium. To detail this process further, we may consider a
nanoparticle consisting of a core in one structural phase covered
by a shell of a different phase. With increasing temperature or
level of excitation, the shell layer's thickness increases and the
optical properties of the particle change continuously from those
of the core phase to those of the surface phase. If the
temperature or level of optical excitation is reduced
\emph{before} the transformation to the new phase is complete,
i.e. while a nucleus of the core phase is still present, the
transformation is reversed, the skin layer shrinks to an
appropriate equilibrium position and the reflectivity returns to
its original level. However, when the core is fully consumed by
the surface phase the particle becomes stable against a return to
the core phase because this would require the creation of a
nucleation center. At this point the applied excitation abruptly
ceases to induce any significant change in the particle's optical
properties, until the temperature approaches the next phase
transition point. This type of optical response, i.e. one based on
reversible, excitation-induced structural transformations, has
been observed previously at bulk gallium/glass interfaces
\cite{Petropoulos01-PRB64} and in gallium nanoparticle films
\cite{MacDonald03-APL82, Pochon04-PRL92} but these experiments did
not resolve the fine structure of transitions involving multiple
phases.

The pump-induced reflectivity change signal observed during the
cooling of the nanoparticulate (see Fig.~\ref{Fig2}c) is very
different from that observed during heating. It is about \emph{two
orders} of magnitude smaller and instead of sharp peaks there are
abrupt step-changes in the signal at certain temperatures. With
the much smaller signal level, there are fewer discernable steps
in the cooling part of the temperature cycle than peaks in the
heating part, and they occur at temperatures different from those
of any of the peaks. Some of these features can be explained by
overcooling: with decreasing temperature the particulate remains
in a given phase until its temperature is somewhat lower than the
normal phase transition temperature and in these conditions pump
excitation produces very little signal. When the overcooled
particulate transforms abruptly into a lower energy phase the
pump-induced probe modulation is again small because this change
happens at a temperature far below the increasing-temperature
signal peak for that phase. Thus, peak $I$ and step $A$ correspond
respectively to transitions into and out of the highest
temperature phase, implying that it overcools by $\sim 45$ $K$;
and step $C$ and peak $V$ are associated with transitions into and
out of the lowest temperature phase, which overcools by $\sim 90$
$K$.

\begin{figure}
\includegraphics{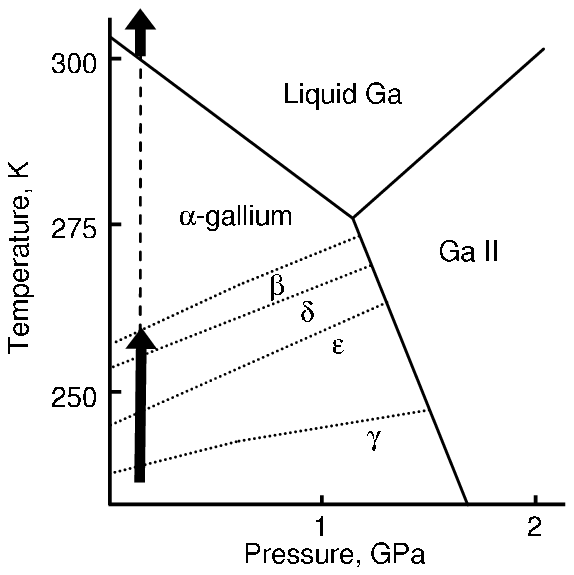}
\caption{Phase diagram for (bulk) gallium after Bosio
\cite{Bosio78-JCP68} showing the sequence of phase transitions
expected in a gallium nanoparticulate undergoing adiabatic heating
from $180$ to $300$ $K$ (The $\alpha$ phase is not found in
nanoparticles so $\beta$ transforms directly to liquid.)
\label{Fig3}}
\end{figure}

Peak $I$ can confidently be attributed to a transition from a
solid state to the liquid. An X-ray diffraction study of gallium
nanoparticles found that $\alpha$-gallium, the stable solid state
of bulk gallium, is completely absent in small particles
\cite{DiCicco98-PRL81} so it follows from Defrain's analysis of
the free energies of gallium's metastable phases
\cite{Defrain77-JCP74} that the only possible sequence of
structural transformations in gallium particles is $\gamma
\rightarrow \varepsilon \rightarrow \delta \rightarrow \beta
\rightarrow liquid$ ($\gamma, \varepsilon, \delta, \beta$ all
crystalline). This sequence is in agreement with the established
phase diagram for gallium shown in Fig.~\ref{Fig3} (the pressure
inside the particulate may be estimated using the Laplace-Young
equation $P = {{4 \sigma} \over {3}} ({2 \over {d}} + {1 \over
{h}})$ where $\sigma$ is the surface tension $\sim 0.7$ $J.m^{-2}$
for the liquid phase, $d = 100$ $nm$ is the particle diameter, and
$h = ft = 18$ $nm$ is its height, giving $P \sim 0.1$ $GPa$).
Thus, peak $I$, at $T_I = 248$ $K$ in Fig.~\ref{Fig2}a, may be
attributed to a transition from monoclinic $\beta$-gallium to the
melt, peak $II$ (at $T_{II} = 246$ $K$) to a transition from
rhombic $\delta$-gallium to $\beta$-gallium and peak $III$ (at
$T_{III} = 237$ $K$) to a transition from $\varepsilon$-gallium
(of unknown point group) to $\delta$-gallium. Assigning
transitions to peak $IV$ (at $T_{IV} = 225$ $K$) and the more
structured peak $V$ (at $T_{V} = 216$ $K$) is more difficult. We
will assume that the $\gamma$ phase, which is strongly represented
in gallium nanoparticle X-ray spectra \cite{DiCicco98-PRL81}, is
the most likely candidate for the ground state (lowest
temperature) phase of the particulate, in which case peak $V$ is
probably associated with a transition from $\gamma$-gallium to
$\varepsilon$-gallium. We believe that peak $IV$ (the smallest of
the five) corresponds to a change in the nanoparticulate's shape
rather than its phase: The optical cross-section of a
nanoparticulate depends strongly on its shape so a particulate's
reflectivity may also change when its shape changes. This argument
is supported by measurements of the retardation between pump and
probe modulations, which relates to the non-instantaneous
relaxation of the high-energy phase back into the low energy phase
following withdrawal of the optical excitation. Small retardation
changes ($\sim$few degrees) are resolved in the vicinity of peaks
$V$ and $III$, and to a lesser extent around peaks $I$ and $II$,
indicating that the relaxation time increases by $5-10$ $\mu s$
close to these peaks. For peak $IV$ the retardation is at least an
order of magnitude larger than for the other peaks - suggesting
that the associated transition is different from the others. In
the absence of any data on the dielectric constants of gallium's
metastable crystalline phases, no information on the phase
transition sequence can be derived from the signs or relative
magnitudes of the peaks in Fig.~\ref{Fig2}a.

Replacing the cw pump beam with a $20$ $kHz$ train of $3$ $ns$
pulses (at the same wavelength) dramatically changes the phase
transition pattern. Such data are presented in Fig.~\ref{Fig2}b,
where the temperature scale has been shifted by $65$ $K$ to
compensate for the strong heating effect induced by the short
pulses and bring the peak pattern back into line with the cw-pump
data in Fig.~\ref{Fig1}a. (Thermodynamic calculations confirm that
the energy absorbed by the nanoparticulate cannot dissipate during
the pulse and therefore rapidly increases its temperature.) For
the pulsed regime of excitation, the optical changes associated
with the phase transitions are very similar in both directions of
temperature, the magnitude of induced reflectivity changes is much
smaller than for cw excitation and there is very little
overcooling. With increasing temperature, peaks $I$ and $V$ are
still present but the large peak at position $III$ is replaced by
a much smoother, broader feature. With decreasing temperature, the
same major peaks are seen with overcooling of only $\sim 2$ $K$
for peak $I$ and $\sim 5$ $K$ for peak $V$. Peaks $II$ and $IV$
from the cw scan cannot be identified. The near-disappearance of
overcooling may result from the fact that a short intense pump
pulse may help a slightly overcooled nanoparticulate to change
from the high temperature phase to the low temperature phase by
rapidly providing sufficient energy for it to get through the
potential barrier between the phases as in the `explosive
crystallization' effect where a localized energy input stimulates
an abrupt transition from an overcooled liquid state to the solid
state \cite{Kuzmenko91-JNCS130}.

The light-induced structural transitions are observed at very low
levels of optical excitation. Such levels can be used because the
differences $\Delta G$ between the free energies of some of the
metastable phases involved are very small: for example, $\Delta
G_{\beta - \delta} = 0.3$ $meV/atom$, $\Delta G_{\gamma - \delta}
= 17$ $meV/atom$ \cite{Defrain77-JCP74}. Thus, the absorbtion of a
`pump' quantum with an energy of $0.8$ $eV$ should be sufficient
to convert about $500$ atoms from the $\gamma$ phase to the
$\delta$ phase, and about $2700$ atoms from the $\delta$ phase to
the $\beta$ phase. The light-induced transitions are likely to be
driven primarily by thermal excitation (i.e. laser-induced
heating) but there may also be a contribution from a
temperature-independent, non-thermal mechanism where the phase
change is caused by band-structure collapse and lattice
instability resulting from electronic excitation
\cite{Guo00-PRL84, Wautelet04-JPCM16}.

To summarize, we have for the first time investigated
light-induced structural transformations in a single gallium
nanoparticulate and observed evidence for a sequence of reversible
transformations between five different structural phases. The low
energy requirements for such optically induced phase transitions,
and the nanoparticulate's phase stability on overcooling could
provide a means of creating key logical and bistable memory
elements for nanophotonic devices operating at extremely low power
levels.

The authors would like to acknowledge the financial support of the
EPSRC (UK) and Funda\c{c}\~{a}o para a Ci\^{e}ncia e Tecnologia
(Portugal).


%



\bibliography{References}

\end{document}